\documentclass[12pt,preprint]{aastex}

\usepackage{graphicx}
\begin{document}

\author{W.A. ${\rm Bruckman}^{1}$, A. ${\rm Ruiz}^{2}$, and E. ${\rm Ramos}^{3}$}
\affil{Department of Physics and ${\rm Electronics}^{1,2}$ and
Department of ${\rm Mathematics}^{3}$\\
University of Puerto Rico at Humacao \\
CUH Station, 100 Route 908, Humacao Puerto Rico 00791-4300}
\email{ ${\tt w\_bruckman@webmail.uprh.edu}^{1}$,${\tt
a\_ruiz@webmail.uprh.edu}^2$,${\tt eramos@www.uprh.edu}^3$}

\title{A Theoretical Model for Mars Crater-Size Frequency Distribution}

\begin{abstract}
We present a theoretical and analytical curve with reproduce
essential features of the frequency distributions vs. diameter, of
the 42,000 crater contained in the Barlow Mars Catalog. The model
is derived using reasonable simple assumptions that allow us to
relate the present craters population with the craters population
at each particular epoch. The model takes into consideration the
reduction of the number of craters as a function of time caused by
their erosion and obliteration, and this provides a simple and
natural explanation for the presence of different slopes in the
empirical log-log plot of number of craters (N) vs. diameter (D).
\end{abstract}

\section{Introduction}

The present impact crater size frequency distribution, N is the
result, on one hand, of a rate of crater formation, $\phi$, and,
on the other hand, the elimination of craters, as time goes by,
due to effects like erosion and obliteration. Therefore if we want
to understand the crater formation history we will need to know
how these forming and erasing factors combine to create $N$. Thus,
in this work the above problem is analyzed, and in section 2 we
find that N can be expressed in terms of $\phi$ and the fractional
reduction of craters per unit of time, C. Then, a simple model is
discussed that describe the crater size distribution in Mars data,
collected by Barlow \citep{bar88}, where it is assumed that $\phi$
is independent of time. The above model is realistic, since
according to several investigations $\phi$ has remained nearly
constant for the last 3 to 3.5 billion years
\citep{har66,neu01,neu83,ryder90}. The simplest interpretation of
this model implies that $\phi$ and $C$ are given as the following
inverse power of the diameter, $D$, of the crater:
$\phi\,\propto\,\frac{1}{D^{4.3}}$,
$C\,\propto\,\frac{1}{D^{2.5}}$. In section 3 the model is applied
to craters data on Earths, and it is concluded that also in our
planet $C\,\propto\,\frac{1}{D^{2.5}}$. This result is interpreted
to mean that on Mars and Earth we have $C\,\sim\,\frac{1}{\rm
Volume}$, or equivalently the crater mean life
$\equiv\,\,\frac{1}{C}\,\propto\,{\rm Volume}\,\propto\,D^{2}\,h$,
with $h\,\propto\,D^{0.5}$. Investigations of geometric properties
of Martian impact craters reflect values of the average height
$h(D)$ consistent with the above conclusion.

\section{Theoretical Models for the Observed Data}

\vspace{0.5cm}

In what follows we will present theoretical and analytical curves
which will reproduce the essential features of the martian
crater-size frequency distribution empirical curves (Figure 1),
based on Barlow's (1987) ${\rm database}$ of about 42,000 impact craters. The
models will be derived using reasonable simple assumptions, that
will allow us to relate the present crater population with the
crater population at each particular epoch.

To this end, let ${\Delta}{N}(D,\tau)$ represents the number of craters of diameter
$D\,{\pm}\,\frac{{\Delta}{D}}{2}$ formed during the epoch $\tau\,{\pm}\,\frac{\Delta{\tau}}{2}$, where
we are assuming that ${\Delta}{D}$ and ${\Delta}{\tau}$ are sufficiently large that is justified
treating ${\Delta{N}}$ as a statistical continuous function, but, on the other hand, they should
be sufficiently small ($\frac{\Delta{\tau}}{\tau}\ll1$,$\frac{\Delta{D}}{D}\ll1$) to be able to treat them as differentials in the following discussion. This initial population will change
as time goes on due to climatic and geological erosion, and the obliteration of
old craters by the formation of new ones. Then, we expect that the change in ${\Delta}{N}$ during
a time interval ${d}{t}$ will be proportional to itself and ${d}{t}$:

\begin{equation}
d({\Delta{N}})\,=\,-C\,{\Delta}{N}{d}{t},
\end{equation}

\noindent where C is the factor that takes into account the
depletion of the craters, and should be a function of the
diameter, since the smaller a crater is the most likely it will
disappear. Furthermore, C could also depend on time however, we
will ignore such changes here, which we believe is a good
starting approximation to the general problem. It is easy to integrate
Equation (1) in time to obtain:

\begin{equation}
{\Delta}{N}(D,t)\,=\,{\Delta}{{N}}(D,t_n)\,{\rm Exp}\left[-C\,{\tau_n}\right],
\end{equation}

\begin{equation}
{\tau}_n\,=\,t - t_n.
\end{equation}

\noindent Equation (2) gives the number of craters, as a function
of $D$, observed at time t, that were produced at the time
interval $t_n\,{\pm}\,\frac{{\Delta}\tau}{2}$. Therefore the total
contribution to the present (t=0) population due to all the epochs
$t_n$ is:

\begin{equation}
N(D)\,=\,\sum_n\,{\Delta}{N}(D,t_n)\,{\rm Exp}\left[  -C{\tau_n}\right],
\end{equation}

\noindent
or in the continuous limit ${\Delta}{\tau}{\rightarrow}{0}$,

\begin{equation}
N(D)\,=\,\int_0^{\tau_f}\,\phi(D,\tau)\,{\rm
Exp}\left[-C{\tau}\right] \,d{\tau},
\end{equation}

\noindent
where

\begin{equation}
\phi(D,\tau)\,\equiv\,{\rm
lim}_{{\Delta}{\tau}{\rightarrow}0}\,\frac{{\Delta}{N{(D,\tau_n)}}}{{\Delta}\tau}.
\end{equation}

\noindent
$\phi(D,\tau)$ is the rate of crater formation of diameter $D$ at the epoch $\tau$, and
$\tau_f$ is the total time of crater formation.

In the next section we will determine the function $C(D)$ and
$\phi$ for a model where we assumed that the rate of crater
formation, $\phi$, is independent of $\tau$.


\begin{figure}
\includegraphics[angle=-90,scale=1.5]{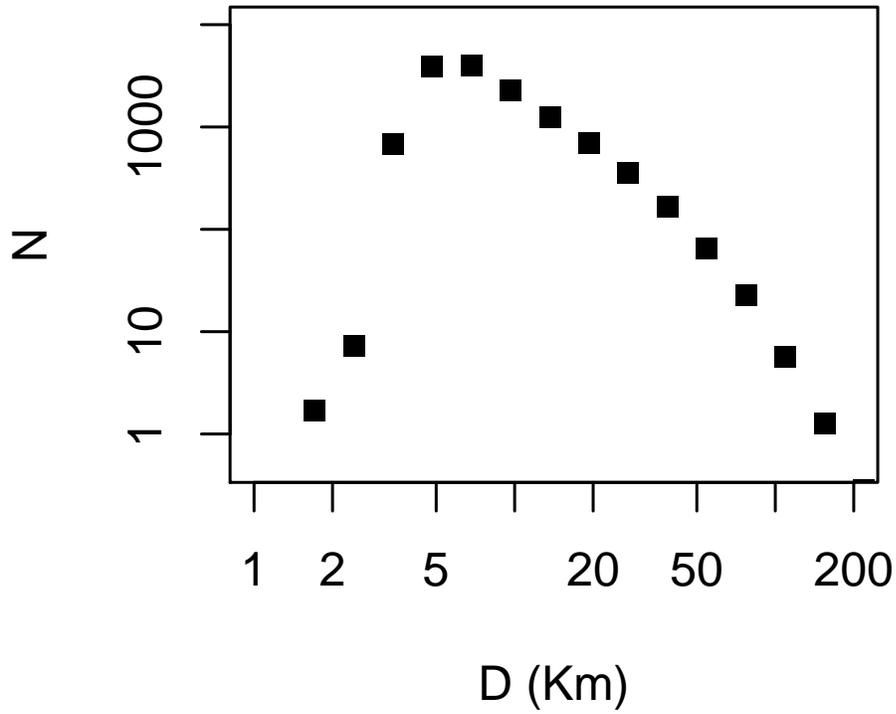}
\caption{Log-log plot of number of craters (N) vs. diameter (D) in
Mars. Following \citep{neu01} the number of craters per kilometer
squared were calculated for craters in the diameter $D_L<D<D_R$,
where $D_L$ and $D_R$ are the left and right bin boundary and the
standard bin width is ${D_R}/{D_L}\,=\,{2}^{1/2}$.}
\end{figure}

\section{$\phi$ Independent of $\tau$}

Investigations of the time dependance of cratering rate of
meteorites have concluded \citep{har66,neu01,neu83,ryder90} that
the impact rate went through a heavy bombardment era that decayed
exponentially until about 3 to 3.5 Gy, and since then has remained
nearly constant until the present. Therefore, for surfaces that
are younger than 3 to 3.5 Gy we can reasonably assume that $\phi$
is independent of $\tau$, and hence from Equation (5) immediately
obtain

\begin{equation}
N(D) = \frac{\phi(D)}{C(D)}\left[1 - Exp\left\{-C\tau_f \right\}\right].
\end{equation}

\noindent We then find that the simplest model that essentially
reproduces the data in Figure 1, for $D\,{\ge}\,6$ km, is given by
Equations (8)and (9):

\vspace{0.5cm}

\begin{equation}
\phi(D)\,=\,\frac{3.55\,{\times}\,10^{9}}{D^{4.3}\,{\tau}_f},
\end{equation}

\begin{equation}
C(D)\,=\,\frac{2.48\,{\times}\,10^{4}}{D^{2.5}\,{\tau}_f}.
\end{equation}

\noindent We see that the theoretical curve (7), shown in
Figure(2), differs significantly from the observed curves for $D$
less than about 6\,km. However, according to Barlow \citep{bar88}
the empirical data is undercounting the actual crater population
for $D$ less than $8$ km, and therefore no meaningful comparison
is then possible between models and data for this region of small
craters.


\begin{figure}
\includegraphics[angle=-90,scale=1.5]{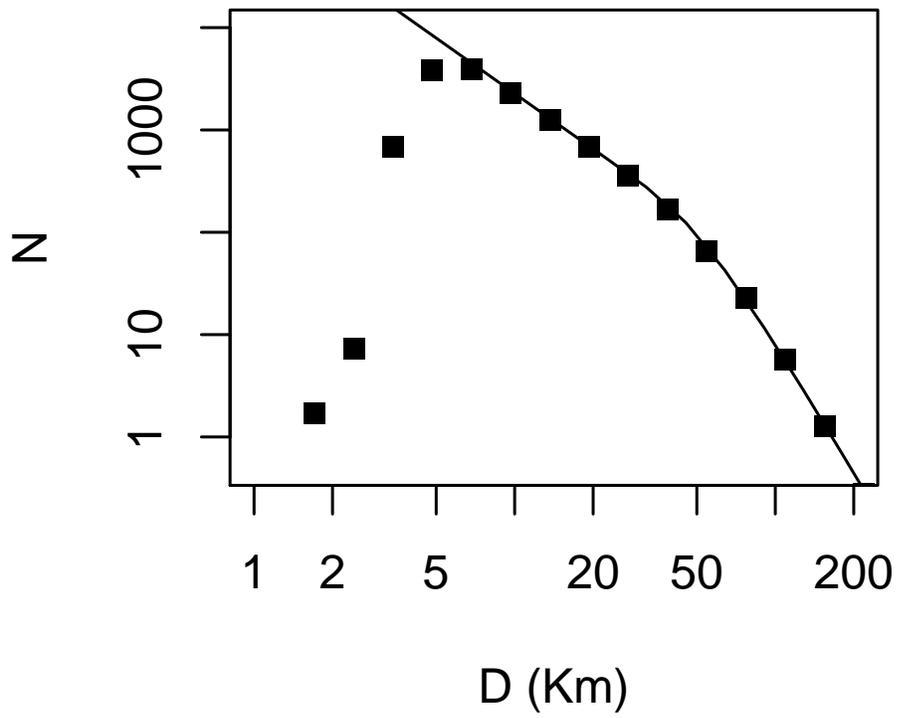}
\caption{Comparison of theoretical model with the empirical
log-log plot of number of craters (N) vs. diameter (D) in Mars.}
\end{figure}

Equation (2) implies that the fraction of craters of diameter $D$
formed at each epoch $\tau$ that still survive at the present time
$\tau=0$ is given by:
\begin{equation}
\frac{{\Delta}N(D,0)}{{\Delta}N(D,\tau)}\,=\,{\rm Exp}\left[-C\tau
\right]\,\approx\,{\rm
Exp}\left[{-\left(\frac{57}{D}\right)}^{2.5}\,\frac{\tau}{\tau_f}
\right]
\end{equation}

\noindent
and thus we have that the mean life for craters of diameter $D$, $\tau_{\rm mean}$, is

\begin{equation}
{\tau}_{\rm mean}\,=\,\frac{1}{C}\,{\approx}\,{\left(\frac{D}{57}\right)}^{2.5}\,{\tau}_{f}.
\end{equation}

\noindent
Hence, craters with $D\,\approx$ 57 km have ${\tau}_{\rm mean}\,\sim\,{\tau}_f$, while

\begin{eqnarray}
\begin{array}{c}
{{\tau}_{\rm mean}\,>>\,{\tau}_{f}\,\,\,\,,\,\,\,\, D\,>>\,57\,{\rm km}}, \\
{{\tau}_{\rm mean}\,<<\,{\tau}_{f}\,\,\,\,,\,\,\,\, D\,<<\,57\,{\rm km}}.
\end{array}
\end{eqnarray}

\noindent
The region $D\,>>\,57$ km is approximately described by the limit of Equation (7) when $D\,\rightarrow\,\infty$:

\begin{equation}
\lim_{D\rightarrow\\{\infty}}\,N\,=\,{\phi}\,\tau_f\,=\,\frac{3.55\,{\times}\,{10}^{9}}{D^{4.3}},
\end{equation}

\noindent which corresponds to a straight line of slope -4.3 in a
Log N vs Log D plot, and that would be the form of Equation (7) in
the absence of erosion and obliterations ($C\,{\approx}\,0$).
Hence,we have that the bending of the empirical curve (Figure 1)
for $D\,{<}\,57\,{\rm km}$ is explained in this model as the
result of the elimination of smaller craters as they get older. We
also see from Equations(13) that when the effect of $C$ can be
ignored we have $N\,=\,\phi\left(D\right)\,\tau_f$, and therefore
the actual crater density $N$ is proportional to the age of the
underlying surface $\tau_f$. On the other hand, when for smaller
craters ${\rm Exp}\left[-C\,\tau_f\right]\,<<\,1$ we will have
from (7) that

\begin{equation}
N(D)\,\approx\,\frac{\phi\left(D\right)}{C\left(D\right)}\,=\,\phi\left(D\right)\tau_{\rm
mean},
\end{equation}

\noindent and in this limit the crater density $N(D)$ is
proportional to the survival mean life, $\tau_{\rm mean}$, of the
craters of size $D$. Thus, when saturation occurs and hence N is
independent of $\tau_f$, we have, instead, that $N$ is
proportional to $\tau_{\rm mean}$. This feature is called by
Hartmann \citep{hart02}``Crater retention age'', and in Mars this
effect shows, according to this model, in craters smaller than
about 57 km.

\section{Application to Earth}

The model given by Equations (7), (8), and (9) assumed a simple
polynomial form for $\phi$ and $C$, however, alternative models
can be also considered. For instance, by assuming that

\begin{equation}
N\,=\,\phi\,\tau_f\,=\,\frac{1.43\,{\times}\,{10}^{5}}{D^{1.8}}\left[1
- {\rm Exp}\left\{
\frac{-2.48\,{\times}\,{10}^{4}}{D^{2.5}}\right\}\right],
\end{equation}

\noindent we will reproduce the Mars crater data, exactly as in
model given by Equations (7),(8),(9) but now with $C\,=\,0$ , and
the change in slope in Figure (1) around $D\,\approx\,57$ km will
now be interpreted as intrinsic behavior of $\phi(D)$ rather than
due to the erosion and obliteration of smaller craters. How can we
then discriminate between these two alternative views?. We see
that in the model given by Equation (7) the fraction of craters of
a given diameter, $D$, produced at a time $\tau$, decreases with
time according to Equation (10) as

\begin{equation}
\frac{{\Delta}N(D,0)}{{\Delta}N(D,\tau)}\,=\,{\rm Exp}\,[-C\tau],
\end{equation}

\noindent while in the model of Equation (15) this fraction is
independent of time. Therefore we can put to test the validity of
Equation (16) by studying crater size frequency distributions as a
function of time. This is possible to do in our planet, and in
this section we will investigate the consistency of the hypothesis
(16) with the Earth craters data.

Thus consider the average diameter of craters observed today that
were formed during a given time $\tau\,\pm\,\frac{d{\tau}}{2}$,
which is given, according to Equation (16), by

\begin{equation}
\bar{D}\,=\,\frac{\int_{0}^{\infty}D\,\phi\,e^{-c\tau}\,d{D}}{\int_{0}^{\infty}\phi\,e^{-c{\tau}}\,d{D}}.
\end{equation}

\noindent Assuming that $C$ and $\phi$ behave in the form

\begin{equation}
\phi\,=\,\frac{A}{D^m}\,,\,\,\,\,A\,=\,{\rm const},\,\,m\,=\,{\rm
const},
\end{equation}

\begin{equation}
C\,=\,\frac{B}{D^l}\,,\,\,\,\,B\,=\,{\rm
const},\,\,l\,=\,{\rm const},
\end{equation}

\noindent we can rewrite Equation (17) in the form (Appendix)

\begin{equation}
\bar{D}\,=\,B^{\frac{1}{l}}\,\alpha\,{\tau}^{\frac{1}{l}},
\end{equation}

\noindent where

\begin{equation}
\alpha\,\equiv\,\,\frac{\Gamma\left(\frac{m-2}{l}\right)}{\Gamma\left(\frac{m-1}{l}
\right)},
\end{equation}

\noindent and

\begin{equation}
\Gamma(n)\,\equiv\,\int_{0}^{\infty}\,U^{n-1}\,e^{-U}\,dU\,
\end{equation}

\noindent is the Gamma function. Equation (20) can be rewritten as

\begin{equation}
{\rm Log}\,{\bar{D}}\,=\,\frac{1}{l}\,{\rm Log}\,\tau + {\rm
Log}\,{\rm B}^\frac{\rm 1}{\rm l}\,\alpha,
\end{equation}


\begin{figure}
\includegraphics[angle=-90,scale=1.5]{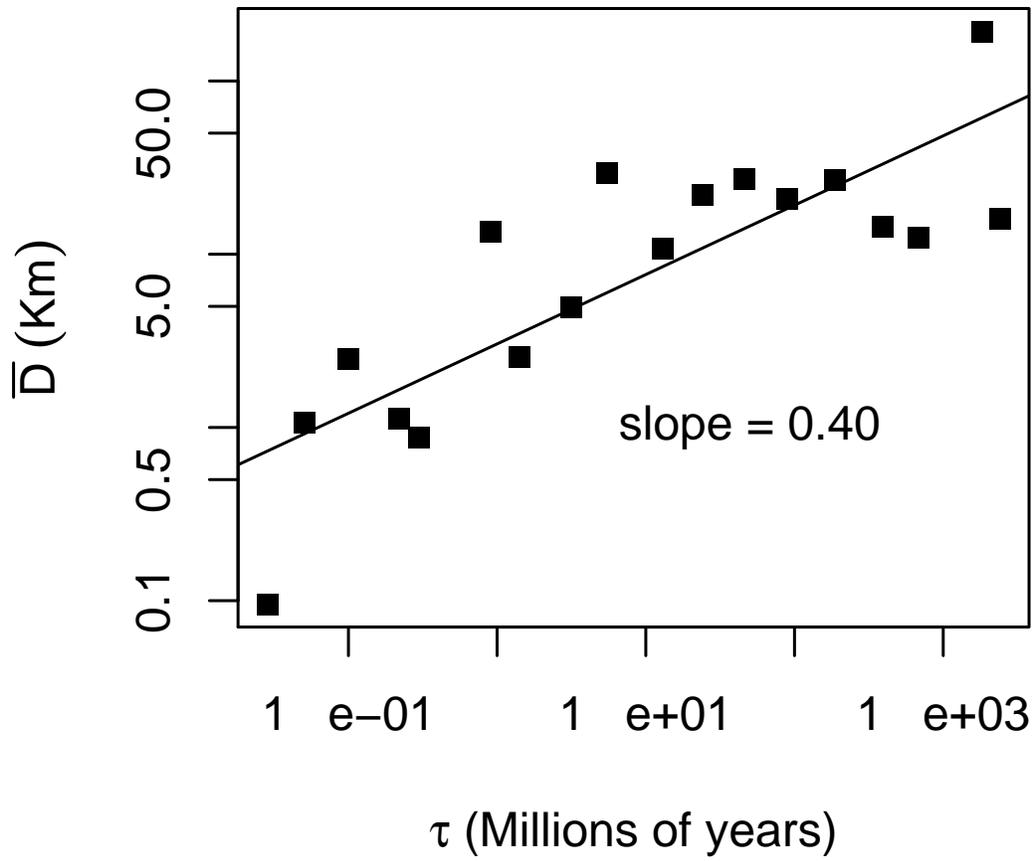}
\caption{Average diameter ($\bar{D}$) vs. average age
($\bar{\tau}$) for terrestrial craters. The bin size increases as
$2^{\frac{n}{2}}$. The slope of the straight line best fit (0.40)
correspond to $l=2.5$.}
\end{figure}

\noindent which represents a linear relation between ${\rm
Log}\,\bar{D}$ and ${\rm Log}\,\tau$ with slope $\frac{1}{l}$. In
Figure (3) we plot ${\rm Log}\,\bar{D}$ vs ${\rm Log}\,\tau$ from
data of crater size vs $\tau$ on Earth, and the straight line best
fitting gives $l\,=\,2.5$, which is the value determined for model
(7) for Mars. This result is interpreted as follows. If we assume
that, as expected, ${\tau}_{\rm mean}$ is a function of the volume
of the crater, $V$, that decreases with decreasing $V$, then it is
reasonable to expand it in terms of powers of ${V}$, and thus we
will have

\begin{equation}
{\tau}\,=\,\frac{1}{C}\,=\,{a_1}{V} + {a_2}{V^2} + {a_3}{V^3} + ...,   .
\end{equation}

\noindent Furthermore, for sufficiently small volumes we would
have, as a good approximation to $C$, that

\begin{equation}
\frac{1}{C}\,{\approx}\,{a}_{1}V\,=\,{a}_{1}\,{D^2}{h},
\end{equation}

\noindent where we are writing

\begin{equation}
V\,=\,D^{2}\,h  ,
\end{equation}

\noindent with $h$ as the average height of the crater of size
$D$. The comparison of Equation (25) with Equations (19), with
$l\,{=}\,2.5$, imply that

\begin{equation}
h\,\sim\,{\rm
Const}{D^\frac{1}{2}},
\end{equation}

\noindent which is a prediction that can be investigated, and we
have found that indeed Equation (27) is consistent with results
from studies of impact crater geometric properties on the surface
of Mars, by J.B. Garrin \citep{gar02}.

Therefore it appears that the age distribution of craters on Earth
favor the simple model considered for Mars, where there is an
erosion and obliteration factor $C$ with the approximate form

\begin{equation}
C\,\approx\,\frac{{\rm Const}}{D^{2.5}}.
\end{equation}

\noindent It is also suggested here that the above behavior for
$C$ follows from a relation of the form

\begin{equation}
C\,\approx\,\frac{{\rm Const}}{V}\,=\,\frac{{\rm Const}}{D^2\,h};
\end{equation}

\noindent with

\begin{equation}
h\,\propto\,D^{\frac{1}{2}}
\end{equation}

Further investigations and observations of the crater data on the
terrestrial planets, the moon and the asteroids are necessary for
additional tests of the validity of the model (7) and its
interpretation.

\section{Appendix}

\noindent
Lets define

\begin{equation}
U\,=\,\frac{B}{D^{l}}\,{\tau},
\end{equation}

\noindent
or, equivalently

\begin{equation}
D\,=\,\left(\frac{B\tau}{U}\right)^{\frac{1}{l}}.
\end{equation}

\noindent
Then we have

\begin{equation}
dD\,=\,\frac{-\left(B\tau\right)^{\frac{1}{l}}dU}{l\,U^{1 + \frac{1}{l}}},
\end{equation}

\noindent and therefore Equation (17) becomes

\begin{equation}
\bar{D}\,=\,\left(B\tau\right)^{\frac{1}{l}}\,\frac{{\int}_{0}^{\infty}U^{\frac{m-2}{l}-1}\,{e}^{-U}{dU}}{\int_0^{\infty}U^{\frac{m-1}{l}-1}\,{e}^{-U}{dU}}\,\equiv\,\left(B\tau\right)^{\frac{1}{l}}\,\alpha,
\end{equation}

\noindent
where $\alpha$ is the ratio of the following gamma functions:

\begin{equation}
\alpha\,\equiv\,\frac{{\int}_0^{\infty}U^{\frac{m-2}{l}-1}e^{-U}{dU}}{{\int}_0^{\infty}U^{\frac{m-1}{p}-1}e^{-U}dU}\,\equiv\,\frac{\Gamma\left(\frac{m-2}{l}\right)}{\Gamma\left(\frac{m-1}{l}\right)}.
\end{equation}

\clearpage

\end{document}